\documentclass[a4paper,11pt]{article}
\usepackage{pos}
\renewcommand{\logo}{\relax}

\title{$\eta$ invariant of massive Wilson Dirac operator and the index}

\author[a]{Shoto Aoki}
\author*[b]{Hidenori Fukaya}
\author[c]{Mikio Furuta}
\author[d]{Shinichiroh Matsuo}
\author[b]{Tetsuya Onogi}
\author[b]{Satoshi Yamaguchi}

\affiliation[a]{
  Graduate School of Arts and Sciences, The University of Tokyo
Komaba, Meguro-ku, Tokyo 153-8902, Japan}

\affiliation[b]{Department of Physics, Osaka University, 
        Toyonaka, Osaka 560-0043 Japan}

\affiliation[c]{Graduate School of Mathematical Sciences, The University of Tokyo, Komaba, Meguro-ku, Tokyo 153-8902, Japan}
\affiliation[d]{Graduate School of Mathematics, Nagoya University, Nagoya, Japan}

\abstract{We revisit the lattice index theorem in the perspective of $K$-theory. The standard definition given by the overlap Dirac operator equals to the $\eta$ invariant of the Wilson Dirac operator with a negative mass. This equality is not coincidental but reflects a mathematically profound significance known as the suspension isomorphism of $K$-groups. Specifically, we identify the Wilson Dirac operator as an element of the $K^1$ group, which is characterized by the $\eta$-invariant.
Furthermore, we prove that, at sufficiently small but finite lattice spacings, this $\eta$-invariant equals to the index of the continuum Dirac operator. Our results indicate that the Ginsparg-Wilson relation and the associated exact chiral symmetry are not essential for understanding gauge field topology in lattice gauge theory.\\

OU-HET-1257
}

\FullConference{The 41st International Symposium on Lattice Field Theory (LATTICE2024)\\
 28 July - 3 August 2024\\
Liverpool, UK\\}

\begin{document}
\maketitle

\section{Introduction}

The index of Dirac operators is a mathematical quantity about the solutions of the Dirac equation of fermions. 
It is defined by the number of zero eigenmodes with the positive chirality and that with the negative chirality.
The Atiyah-Singer index theorem \cite{Atiyah:1968mp}
shows that this quantity equals to the topological charge
of the background gauge fields.
The index has been an important subject both in physics and mathematics
to understand the gauge field topology, which is nonperturbative.

In lattice gauge theory, both of the Dirac index and topological charge of the gauge fields
are difficult to describe.
It is difficult to identify the chiral zero modes with the standard lattice Dirac operators which
break the chiral symmetry \cite{Nielsen:1980rz,Nielsen:1981hk,Nielsen:1981xu}.
Lattice discretization of spacetime makes the notion of topology obscure.

A traditional solution was given by the overlap Dirac operator \cite{Neuberger:1997fp} 
(and \cite{Hasenfratz:1998ri}).
With the overlap Dirac operator satisfying the Ginsparg-Wilson relation \cite{Ginsparg:1981bj},
one can define an exact modified chiral symmetry on the lattice \cite{Luscher:1998pqa}
and the index given by the trace of the modified chirality operator is well-defined.
However, this solution has been so far limited to even-dimensional periodic square lattices
whose continuum limit is a flat torus.

The overlap Dirac operator with the lattice spacing $a$ is given by
\begin{equation}
  \label{eq:ovdef}
D_{\rm ov} = \frac{1}{a}\left(1+\gamma_5{\rm sgn}(H_W)\right),
\end{equation}
where $H_W=\gamma_5(D_W-M)$ is the Wilson Dirac operator \cite{Wilson1977}
with a negative mass we often take $M=1/a$.
This operator satisfies the Ginsparg-Wilson relation,
\begin{equation}
\gamma_5 D_{\rm ov}+D_{\rm ov}\gamma_5= aD_{\rm ov}\gamma_5D_{\rm ov},
\end{equation}
with which the fermion action $S = \sum_x \bar{q}(x) D_{\rm ov}q(x) $
is invariant under the ``modified'' chiral rotation:
\begin{equation}
q \to e^{i\alpha\gamma_5(1-aD_{\rm ov})}q,\;\;\;\bar{q}  \to \bar{q}e^{i\alpha\gamma_5}.
\end{equation}
Moreover, it reproduces the axial $U(1)$ anomaly from the
fermion measure,
\begin{equation}
Dq\bar{q} \to \exp\left[2i\alpha {\rm Tr}(\gamma_5+\gamma_5(1-aD_{\rm ov}))/2\right]D q\bar{q}.
\end{equation}
The index of the overlap Dirac operator is defined as
\begin{equation}
{\rm Ind}D_{\rm ov} = {\rm Tr}\gamma_5\left(1-\frac{aD_{\rm ov}}{2}\right),
\end{equation}
and it equals to
the difference of the number of zero modes with the $\pm$ chiralities.

One should remember, however, that $D_{\rm ov}$ is a function of the
Wilson Dirac operator.
Substituting Eq.~(\ref{eq:ovdef}) into the definition of the index,
a nontrivial equality is obtained:
\begin{equation}
{\rm Ind}D_{\rm ov} = -\frac{1}{2}{\rm Tr}\;{\rm sgn}(H_W) =: -\frac{1}{2} \eta(H_W),
\end{equation}
where $\eta(H_W)$ is known as the $\eta$ invariant in mathematics, which was introduced by Atiyah, Patodi and Singer \cite{Atiyah:1975jf}.
In this talk, we show a $K$-theoretic meaning of the right-hand side of the equality,
and try to convince the readers that the massive Wilson Dirac operator is an equally good or even better mathematical object
in terms of $K$-theory \cite{MR0233870,MR0285033}
than the overlap Dirac operator to describe the gauge field topology\footnote{
Recently there are some developments in mathematics.
The Atiyah-Singer index theorem was directly formulated in \cite{Yamashita:2020nkf} on a lattice
using the algebraic index theorem of Nest and Tsygan.
In \cite{Kubota:2020tpr}, the lattice approximation of analytic indices through
the higher index theory of almost flat vector bundles was formulated.
}.

We notice that this work is the first lattice version
of the phys-math project on the ``physicist-friendly'' index
of the Dirac operators \cite{Fukaya:2017tsq,Fukaya:2019qlf,Fukaya:2020tjk,Fukaya:2021sea}, 
in which we are reformulating the index in terms of massive (domain-wall)
fermions without imposing any nonlocal boundary conditions.
We expect a wider application of our new formulation
than the standard definition by the overlap Dirac operator
since the extension to the systems with boundaries
and those with higher symmetries using $KO$ or $KSp$ groups
is straightforward.
A mathematical proof of this work was already
given in Ref.~\cite{Aoki:2024sjc}.

\section{$K$-theory}
$K$-theory \cite{MR0233870,MR0285033} is one of the generalized cohomology theories
and useful in classifying the vector bundles.
Here we briefly describe the essence of $K$-theory
and how it describes the index of the Dirac operators.

A vector bundle is a united manifold that consists of a base space $X$
and a fiber space $F$ which is a vector space.
In physics, a field $\psi(x)$ at a position $x$ can be identified
a position $(x,\psi)$ in the vector bundle.
The vector bundle is not a direct product $X\times F$ but
twisted by gauge fields or connections.
The total space is often denoted by $E$.

The element of $K^0(X)$ group is given by a set $[E_1, E_2]$
where $E_1$ and $E_2$ are two vector bundles over the same base space $X$.
Equivalently, we can consider an operator and its conjugate\footnote{
To be precise, the operators act on the sections of $E_i$.
},
\begin{equation}
D_{12}: E_1\to E_2,\;\;D^\dagger_{12}: E_2\to E_1,
\end{equation}
to represent the same element by $[E,D,\gamma]$, where
\begin{equation}
E=E_1\oplus E_2, \;\;D=\left(
\begin{array}{cc}
 & D_{12}\\
D^\dagger_{12} &
\end{array}
\right),\;\; \gamma=\left(
\begin{array}{cc}
1 & \\
& -1
\end{array}
\right).
\end{equation}  
If we identify $E_1$ as the vector space of the left-handed spinor,
and $E_2$ as that of the right-handed,
the Dirac operator $D$ can be identified as the $K^0(X)$ group element.
Namely, the $K^0(X)$ group classifies the massless Dirac operator
which anticommutes with the chirality operator $\gamma$.

When we are only interested in a global structure of the gauge fields,
we can forget about details of the base manifold $X$ taking 
a ``one-point compactification'' by the so-called $K$-theory push-forward $G : K^0(X)\to K^0(\mbox{point})$: 
\begin{equation}
[E,D,\gamma]\mapsto[\mathcal{H}_E, D, \gamma],
\end{equation}
where $\mathcal{H}_E$ is the whole Hilbert space on which $D$ acts\footnote{
This map is analogous to the forgetful map from a complex number
to two real numbers $f:\mathbb{C}\to \mathbb{R}^2$ where the map is given by $a+ib \to (a,b)$
or the type conversion in program codes from complicated one to simple one like from ``str'' to ``list''.
}.
Many information is lost with the map, but one (the Dirac operator index) remains.

In this work, we would like to suspend the ``point''
to an interval $I$ with two ends denoted by $\partial I$.
In $K$-theory, there is an important isomorphism called the suspension isomorphism
$K^0(\mbox{point}) \cong K^{1}(I, \partial I )$ with the one-to-one map
\begin{equation}
[\mathcal{H}_E, D, \gamma] \leftrightarrow [p^*\mathcal{H}_E, D_t]
\end{equation}
exists where the one-parameter family of the Dirac operator $D_t$ is labeled by $t \in I=[-1,1]$
with a condition that $D_{\pm 1}$ is invertible.
$p^*$ denotes the pullback of the projection $p:I\to \mbox{point}$.
In the following analysis, we take $I$ as a parameter space of the fermion mass,
and the same Hilbert space $\mathcal{H}_E$ is extended along $I$ in a trivial way.
We, therefore, omit $p^*$ for simplicity in the following.
The physical meaning of the isomorphism will be given soon later.

\section{Massless Dirac vs. massive Dirac}

In the standard formulation of the Dirac operator index,
we need a massless Dirac operator and
its zero modes with definite chirality.
Therefore, we need to consider the $K^0(\mbox{point})$ elements
given by $[\mathcal{H}_E, D, \gamma]$.
But we will show that it is isomorphic to
$K^{1}(I, \partial I )$ whose element is described
by the massive Dirac operators $[\mathcal{H}_E, \gamma(D+m)]$
where the mass term is varied in the range $-M\le m \le +M$.

Let us consider the eigenvalues of the continuum massive Dirac operator $H(m):=\gamma(D+m)$.
For the original zero mode in the massless case, it is still an eigenmode
with the eigenvalue $\pm m$ and the sign is equal to the chirality.
For the nonzero modes, the anticommutation relation
$[D,H(m)]$ guarantees pairings of the eigenvalues $\pm\sqrt{\lambda_0^2+m^2}$,
where the $\lambda_0$ is the nonzero eigenvalue of $H(m=0)=\gamma D$.

\begin{figure*}[tbhp]
  \centering
  \includegraphics[width=0.5\columnwidth]{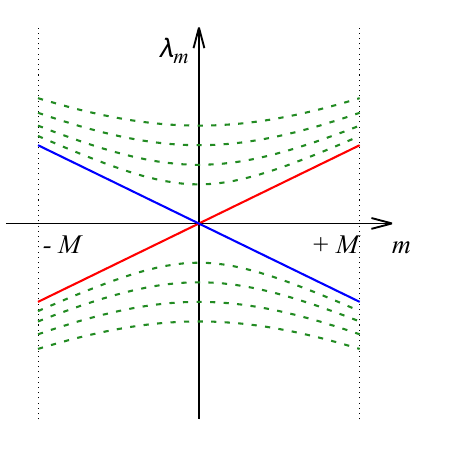}
  \caption{
    Eigenvalue spectrum of the massive Dirac operator $H(m)$ as a function of $m$.
  }
  \label{fig:chiralEigen}
\end{figure*}

Let us illustrate in Fig.~\ref{fig:chiralEigen}
the spectrum of this massive Dirac operator as a function
of mass parameter $m \in [-M,M]$.
The nonzero modes
make parabolic (dashed-green) curves
which are $\pm$ symmetric.
On the other hand, the chiral modes cross zero from negative to positive
when its chirality is $+$ (red-solid line) and go from positive to negative
when they have the negative chirality (blue-solid).

If we count the net zero-crossing lines or the so-called spectral flow,
the Atiyah-Singer index is obtained. 
Note that whenever a eigenvalues crosses zero, the $\eta(H(m))$ jumps by two, 
and therefore, it is also equal to the difference of the $\eta$ invariants,
\begin{equation}
  {\rm Ind} D = \frac{1}{2}\left[\eta(H(M))-\eta(H(-M))\right].
\end{equation}

This describes the physical meaning of the suspension isomorphism
$K^0(\mbox{point}) \cong K^{1}(I, \partial I )$.
The standard definition of the Dirac operator index
which characterizes $[\mathcal{H}_E,D,\gamma]$ is always equal to
the spectral flow along the one parameter family $-M\le m\le +M$
of the massive Dirac operator $\gamma(D+m)$ which
characterizes $[p^*\mathcal{H}_E, \gamma(D+m)]$.
The massless definition corresponds to counting the index by points
at $m=0$, while the massive case counts the same index by lines crossing zero
in the range $[-M,M]$.
The two definitions of the index always agree.

The equivalence between the index to
the spectral flow of the Wilson Dirac operator was known in physics.
It was found in a rather empirical way at the early stage \cite{Itoh:1987iy}
but later a mathematically rigorous equivalence was 
established \cite{Adams:1998eg}. 
But as far as we know,
the mathematical relevance of the Wilson Dirac operator as the element of
the $K$ group
has not been discussed. 

\begin{figure*}[tbhp]
  \centering
  \includegraphics[width=0.5\columnwidth]{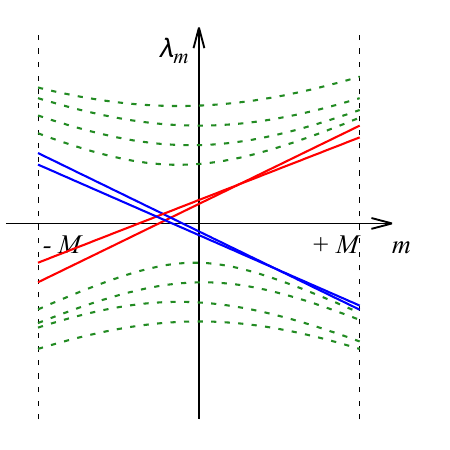}
  \caption{
    Eigenvalue spectrum of the massive Dirac operator $H(m)$
    with chiral symmetry breaking effect, modeling
    the situation with a lattice Dirac operator.
  }
  \label{fig:nonchiralEigen}
\end{figure*}

What happens when a chiral symmetry breaking regularization
is employed (on a lattice)?
The spectrum will be deformed like Fig.~\ref{fig:nonchiralEigen}.
The nonzero modes are no more $\pm$ symmetric
and the chiral modes do not cross zero at $m=0$.
It is difficult to identify the chiral zero modes to
give the standard definition of the Dirac operator index.
However, we can still count the zero-crossing  lines
as far as the two end points at $m=\pm M$ have a enough gap
from zero in the spectrum.
Counting lines is easier and stabler against the symmetry breaking
than counting points.
This is the core of this work and explains why
the Wilson Dirac operator works as the $K^1(I,\partial I)$
group element to describe the index on the lattice.
The $K^1(I,\partial I)$ group is insensitive to the existence of
the chirality operator $\gamma$ and easier to consider
in lattice gauge theory.

\section{Main theorem}

In this section, we briefly sketch our main theorem in a physicist-friendly way.
See the full mathematical proof given in Ref.~\cite{Aoki:2024sjc}.

We consider a complex vector bundle $E$ over a $2n$-dimensional flat continuous torus $T^{2n}$
with an integer $n$. 
We give a connection to this bundle by a gauge field
whose definition is given below.
Let us denote the standard continuum Dirac operator which acts on
a section of $E$ by
\begin{equation}
D = \gamma^\mu(\partial_\mu + A_\mu),
\end{equation}
where the Dirac matrices satisfy $\{\gamma_\mu,\gamma_\nu\}=2\delta_{\mu\nu}$
and the chirality operator or $\mathbb{Z}_2$ grading operator is
given by $\gamma=i^n \prod_\mu \gamma_\mu$.
Note that the Hilbert space $\mathcal{H}_E$ to which $D$ operates is infinite-dimensional
but the dimension of zeromode subspace, as well as that of $D^\dagger$ are finite in general.
Such $D$ is called Fredholm.

We regularize the torus $T^{2n}$ by a square lattice with a lattice spacing $a$.
Note that the fiber vector space is kept continuous. 
There is some ambiguity in defining the link variables from continuum theory.
The standard approach is to define the continuum gauge field first
and then to define the link variables by the Wilson line, 
taking the path-ordered product of the exponentiated gauge field connection.
In this work, we take an opposite direction to simplify the analysis.
We first define the link variables between arbitrary two points $(x,y)$ on the torus:
\begin{equation}
 U(x,y) \in {\rm Hom}(E_x,E_y),
\end{equation}
where $E_{x,y}$ is the restriction of the bundle onto $x,y$.
We assume that $U(x,y)$ satisfies the conditions $U(x,x)=1$ and $U(y,x)=U(x,y)^{-1}$.
Then the link variables at arbitrary lattice spacing are uniquely defined by $U(x,x+a e_\mu )$,
where $e_\mu$ is a unit vector in the $\mu$-direction of $T^{2n}$.
The continuum gauge field is also uniquely determined up to gauge transformation by
\begin{equation}
 A_\mu(x) = -\lim_{\epsilon\to 0}\frac{\varphi^{-1}(x)U(x,x+\epsilon e_\mu)\varphi(x+\epsilon e_\mu)}{\epsilon},
\end{equation}
where $\varphi$ is a local trivialization on an open patch containing both points $x$ and $y$.

The Wilson Dirac operator is given in the standard way
in terms of the above link variables.
\begin{equation}
 D_W = \sum_{\mu=1}^{2n} \left[\gamma^\mu \frac{\nabla_\mu-\nabla_\mu^*}{2}+\frac{a}{2}\nabla_\mu\nabla_\mu^*\right].
\end{equation}
where $\nabla_\mu$ is the forward covariant difference operator acting as $\nabla_\mu\psi(x)=[U(x,x+a e_\mu)\psi(x+a e_\mu)-\psi(x)]/a$,
and $\nabla_\mu^*$ is its conjugate.
Here we have chosen the Wilson coefficient unity.

To define the $K^1(I,\partial I)$ group,
we consider Hilbert bundles taking the base space as
$I=[-M,M]$ which is parametrized by the mass $m$
and the fiber space $\mathcal{H}$ as the one-parameter family of
the Hilbert space to which the Dirac operator $D_m$ operates.
The group elements are given by the equivalence classes of 
$[\mathcal{H}, D_m]$ having the same spectral flow.
Note again that the $K^1(I,\partial I)$ group does not require
any chirality operator.

The group operation is given by taking the direct sum,
\begin{equation}
 [(\mathcal{H}^1,D^1_m)\} \pm \{(\mathcal{H}^2,D^2_m)] =  [(\mathcal{H}^1\oplus \mathcal{H}^2,\left(\begin{array}{cc}D^1_m &\\& \pm D^2_m\end{array}\right))]
\end{equation}
and the identity element is defined by
\begin{equation}
 [(\mathcal{H},D_m)]|_\text{Spec.flow=0}.
\end{equation}
Note that in this definition, the dimension of the Hilbert spaces
can be either finite or infinite so that the continuum and lattice
Dirac operators can be equally treated.

We compare the continuum and lattice Dirac operators
treating them as  the $K^1(I,\partial I)$ group elements,
$[(\mathcal{H}_\text{cont.},\gamma(D_{\rm cont.}+m))]$
and $[(\mathcal{H}_\text{lat.},\gamma(D_W+m))]$, respectively.
It is sufficient to consider a continuum-lattice combined Dirac operator
\begin{equation}
\hat{D}=\left(\begin{array}{cc}\gamma(D_{\rm cont.}+m) & tf_a\\tf_a^* & - \gamma(D_W+m)\end{array}\right),
\end{equation}
which is given by the difference
$[(\mathcal{H}_\text{cont.},\gamma(D_{\rm cont.}+m))]-[(\mathcal{H}_\text{lat.},\gamma(D_W+m))]$
and adding a perturbation: the continuum-lattice mixing mass term $tf_a$ and its conjugate.
Here, $f_a$ is a map from $\mathcal{H}_\text{lat.}$ to $\mathcal{H}_\text{cont.}$,
interpolating the vectors on the discrete lattice sites 
to those on the continuum position space by a product of 
simple linear partition of unity in every direction.
The conjugate $f_a^*$ corresponds to the coarse graining map from
the continuum Hilbert space to the discrete space on the lattice.

We prove by contradiction for any gauge field background determined by $\{U(x,y)\}$
that there exists a finite lattice spacing $a_0$,
such that for any lattice spacing $a<a_0$, 
$\hat{D}$ is always invertible, having no zero mode,
along the path in the two parameter space $(m,t)=(-M,0)\to (-M,1)\to (+M,1)\to (+M,0)$.
This is a sufficient condition for $[(\mathcal{H}_\text{cont.},\gamma(D_{\rm cont.}+m))]$
and $[(\mathcal{H}_\text{lat.},\gamma(D_W+m))]$ to be
in the same equivalence class.
We can show that the $\eta(\gamma(D_W+M))=0$ for $M>0$
and therefore, we obtain
\begin{equation}
 {\rm Ind}D_{\rm cont.} = -\frac{1}{2}\eta(\gamma(D_W-M)),
\end{equation}
for $a<a_0$.

\section{Comparison with the overlap Dirac index}

So far we have shown that the
Wilson Dirac operator is as good as the overlap Dirac operator $D_{\rm ov}$
to describe the index proving using $K$-theory that
the $\eta$ invariant $\eta(\gamma(D_W-M))$ and corresponding 
continuum one agree for sufficiently small lattice spacings.
In this section, we will show that
$D_W$ has a potentially wider application than $D_{\rm ov}$
to the following systems.

One possible application is the index on manifolds with boundaries.
In the previous sections, we have only considered a 
periodic square lattice, whose continuum limit is a flat torus.
For the massive Dirac operators, it is straightforward
to introduce a free boundary condition, 
to obtain the domain-wall fermion Dirac operator $D_{\rm DW}$ \cite{Kaplan:1992bt, Furman:1994ky,Shamir:1993zy}.
In \cite{Fukaya:2017tsq,Fukaya:2019qlf}
we have shown in continuum theory that 
the $\eta$ invariant of domain-wall Dirac operator
is equal to the Atiyah-Patodi-Singer(APS) index
of the Dirac operator with a nontrivial boundary.
Therefore, it is natural to conjecture that
$-\frac{1}{2}\eta(\gamma(D_{\rm DW}))$ at sufficiently
small lattice spacings agrees with the APS index.
The overlap Dirac operator, in contrast, 
is no more chiral symmetric when we impose such a boundary condition,
since the boundary condition invalidates the Ginsparg-Wilson relation \cite{Luscher:2006df}.

Another application is the case of real Dirac operators.
The Dirac operator is real when there exists a symmetric real operator $C$
which satisfies $D^*=CDC^{-1}$.
A famous example is the Dirac operator in five-dimensional $SU(2)$ theory
in the fundamental representation, which is the source of the 
Witten anomaly \cite{Witten:1982fp} in four dimensional chiral gauge theory.
For general complex Dirac operators, the $K^1(I,\partial I)$ group 
is characterized by the $\eta$ invariant.
For real Dirac operators, we have $KO^0(I,\partial I)$ group, 
which is characterized by the mod-two spectral flow given by
\begin{equation}
 -\frac{1}{2}\left[1-{\rm sgn}\det\left(\frac{D-M}{D+M}\right)\right]
\end{equation}
and the mod-two Atiyah-Singer index by the suspension isomorphism to $K^1({\rm point})$.
It is then natural to conjecture that the lattice version given by the standard
Wilson Dirac operator
$-\frac{1}{2}\left[1-{\rm sgn}\det\left(\frac{D_W-M}{D_W+M}\right)\right]$
at sufficiently small lattice spacings agrees with the mod-two index.
In contrast, it is unknown how to construct the overlap-like operator 
in such a case.

We also expect that a nontrivial gravitational background can 
be implemented by a curved domain-wall mass term 
\cite{Aoki:2022cwg,Aoki:2022aez,Kaplan:2023pxd,Kaplan:2023pvd,Aoki:2024bwx,Aoki:2024bwx,Aoki:2024nvk,Kaplan:2024ezz}.
It would be interesting to formulate, for instance,
the $\hat{A}$ genus of the curved domain-wall on a lattice.

\section{Summary}

In this work, we have shown that the massive Wilson Dirac operator is an equally good or even better
object than the overlap Dirac operator to describe the gauge field topology.
The chiral symmetry is not essential and massive operators are good enough
to be identified as the $K^1$ group elements.
We have proved by contradiction that the spectral flow, or equivalently 
the $\eta$ invariant, of the massive Wilson Dirac operator agrees
with that in the continuum theory at sufficiently small but finite lattice spacings.
Its equality to the standard definition of the continuum Dirac index
is guaranteed by the suspension isomorphism between $K^0({\rm point})$ and $K^1(I,\partial I)$ groups.

In terms of $K$-theory, we expect wider application of our formulation
than the standard overlap Dirac operator index,
to more nontrivial systems where the chiral symmetry is difficult or absent,
such as the case when the Dirac operator is real,
the one with nontrivial boundaries, curved gravitational background, and so on.

We thank Mayuko Yamashita for her significant contribution at the
early stage of this work. We thank Sinya Aoki, Yoshio Kikukawa, Yosuke Kubota, and Hersh Singh for useful
discussions. This work was partly supported by JSPS KAKENHI Grant Numbers JP21K03222,
JP21K03574, JP22H01219, JP23K03387, JP23K22490, JP23KJ1459.

\bibliographystyle{JHEP}
\bibliography{references.bib}



\end{document}